\documentclass[final]{IEEEtran}

\usepackage{cite}
\usepackage{color}
\usepackage{threeparttable}
\usepackage[pdftex]{graphicx}
\usepackage{epstopdf}
\usepackage{picinpar}
\usepackage[cmex10]{amsmath}
\usepackage{amsmath,amsfonts,amssymb}
\usepackage{subfigure}
\usepackage{changepage}
\usepackage{algorithm}
\usepackage{caption}
\usepackage{algpseudocode}
\usepackage{stfloats}
\usepackage{bm}
\usepackage{amsthm}
\usepackage{enumerate}
\usepackage{multirow}
\usepackage{booktabs}
\usepackage{diagbox}
\usepackage{balance}
\usepackage[colorlinks,linkcolor=red, citecolor=blue, urlcolor=black]{hyperref}
\usepackage{graphicx}

\allowdisplaybreaks

\begin{document}

\title{{LUNA: Luneburg-Lens-Aided Reconfigurable Array for 6G-and-Advanced Wireless Networks}}

\author{
Ziwei Wan,~\IEEEmembership{Member,~IEEE},
Zhen Gao,~\IEEEmembership{Senior Member,~IEEE},
Shuping Dang,~\IEEEmembership{Senior Member,~IEEE},
Michail Matthaiou,~\IEEEmembership{Fellow,~IEEE},
Zhaocheng Wang,~\IEEEmembership{Fellow,~IEEE},
and Sheng Chen,~\IEEEmembership{Life Fellow,~IEEE}

\thanks{Z. Wan and Z. Gao are with the School of Interdisciplinary Science, Beijing Institute of Technology (BIT), Beijing 100081, China (e-mails: \{ziweiwan,gaozhen16\}@bit.edu.cn).}

\thanks{Shuping Dang is with School of Electrical and Electronic Engineering, Nanyang Technological University, Singapore 639798 (e-mail: shuping.dang@ntu.edu.sg).}

\thanks{M. Matthaiou is with the Centre for Wireless Innovation (CWI), Queen’s University Belfast, BT3 9DT Belfast, U.K. (e-mail: m.matthaiou@qub.ac.uk).}

\thanks{Z. Wang is with the Department of Electronic Engineering, Tsinghua University, Beijing 100084, China (e-mail: zcwang@tsinghua.edu.cn).}

\thanks{S. Chen is with the School of Electronics and Computer Science, University of Southampton, SO17 1BJ Southampton, U.K. (e-mail: sqc@ecs.soton.ac.uk).}
}

\maketitle

\begin{abstract}

This article introduces the {{LU}neburg-lens-aided reco{N}figurable {A}rray} (LUNA), an antenna architecture that unifies multiple-input multiple-output (MIMO) and network-controlled repeater (NCR) functionalities in the Luneburg lens-enabled hardware platform.
A Luneburg lens, fabricated from graded-index dielectric materials, passively converts the radiation of a low-gain feed into a highly directional beam without active phase shifting, while a dense passive feed bank and a reconfigurable feed-selection network electronically switch the beam directions with minimal hardware complexity and power consumption.
We commence by reviewing the basic principles and application history of Luneburg lenses in radar and wireless communications, which motivates their role in 6G-and-advanced networks.
Then, we highlight how a Luneburg lens and a reconfigurable feed array construct both LUNA-MIMO and LUNA-NCR, where the lens and feed bank can be reused across functions and frequency bands.
Case studies demonstrate that LUNA achieves the satisfactory spectral and energy efficiency with a few radio-frequency chains, and it also improves positioning performance for sensing tasks.
Finally, some open problems and research directions are provided to inspire follow-up research on LUNA.

\end{abstract}

\section{Introduction}\label{S1}

As wireless networks evolve from current 5G toward 6G and beyond, they are expected to transcend high-throughput connectivity and evolve into a deeply integrated infrastructure that simultaneously supports ubiquitous communications, high-precision sensing, intelligent interaction, and resilient coverage \cite{itu2023framwork}. 
Realizing this vision hinges critically on the antenna front end, which must deliver high beamforming gain, broad spatial adaptability, high energy efficiency, and scalable hardware at a time when conventional multiple-input multiple-output (MIMO) architectures are approaching fundamental limits in cost, power, and complexity.
The recently emerged movable antenna (MA) concept \cite{zhu2026tutorial} has attracted growing interest for providing additional spatial degrees of freedom (DoFs) without scaling up array size. Although a small number of actively radiating elements can be physically repositioned to exploit channel spatial variations in MAs, the resulting aperture is often insufficient to overcome the severe propagation loss, especially in the millimeter-wave and terahertz bands. This fundamental trade-off between the DoFs of motion and the aperture required for array gain motivates alternative architectures that decouple beamforming capability from the number of active antennas.

As a remedy, this article introduces the {{\bf LU}neburg-lens-aided reco{\bf N}figurable {\bf A}rray} (LUNA), which couples the electromagnetic focusing capability of Luneburg lenses \cite{Original} with electronically reconfigurable feeds.
A Luneburg lens is a dielectric sphere whose refractive index decreases from the center to the surface. It can passively transform feed radiation into highly directional beams, and, by reciprocity, focus an incoming plane wave onto a distinct focal point determined by its angle of arrival.
This angular-to-spatial mapping is one-to-one and frequency-invariant due to the spherical symmetry, and therefore a single Luneburg lens together with a switchable feed bank can serve diverse frequency bands without replacing the whole front end.
More importantly, LUNA makes beamforming capability built not on power-hungry active phase shifters but on a passive lens, while a small number of radio-frequency chains (RFCs) handle only the selected feeds to achieve nearly-optimal performance.

This article envisions how the LUNA paradigm will reshape future 6G networks.
The key features of this article are threefold:
\begin{itemize}
{\item An overview of Luneburg lenses is given, briefly introducing the history, operating principle, implementations, and applications. We demonstrate that the success of Luneburg lenses in both radar and communications paves the way for their application in future wireless networks.
}
{\item We introduce LUNA as a versatile architecture supporting both MIMO and network-controlled repeater (NCR) operations. We also present the corresponding transceiver structures, and our case studies showcase the potential superiority of LUNA over conventional array schemes.
}
{\item Open problems and future research directions for LUNA-aided networks are discussed, including theoretical analysis, hardware design, and synergies with artificial intelligence (AI).
}
\end{itemize}

\section{Preliminaries of Luneburg Lens}\label{S2}

\begin{figure*}[!tp]
\captionsetup{singlelinecheck = off, font={footnotesize}, name = {Fig.}, labelsep = period}
\centering
\includegraphics[width=2\columnwidth]{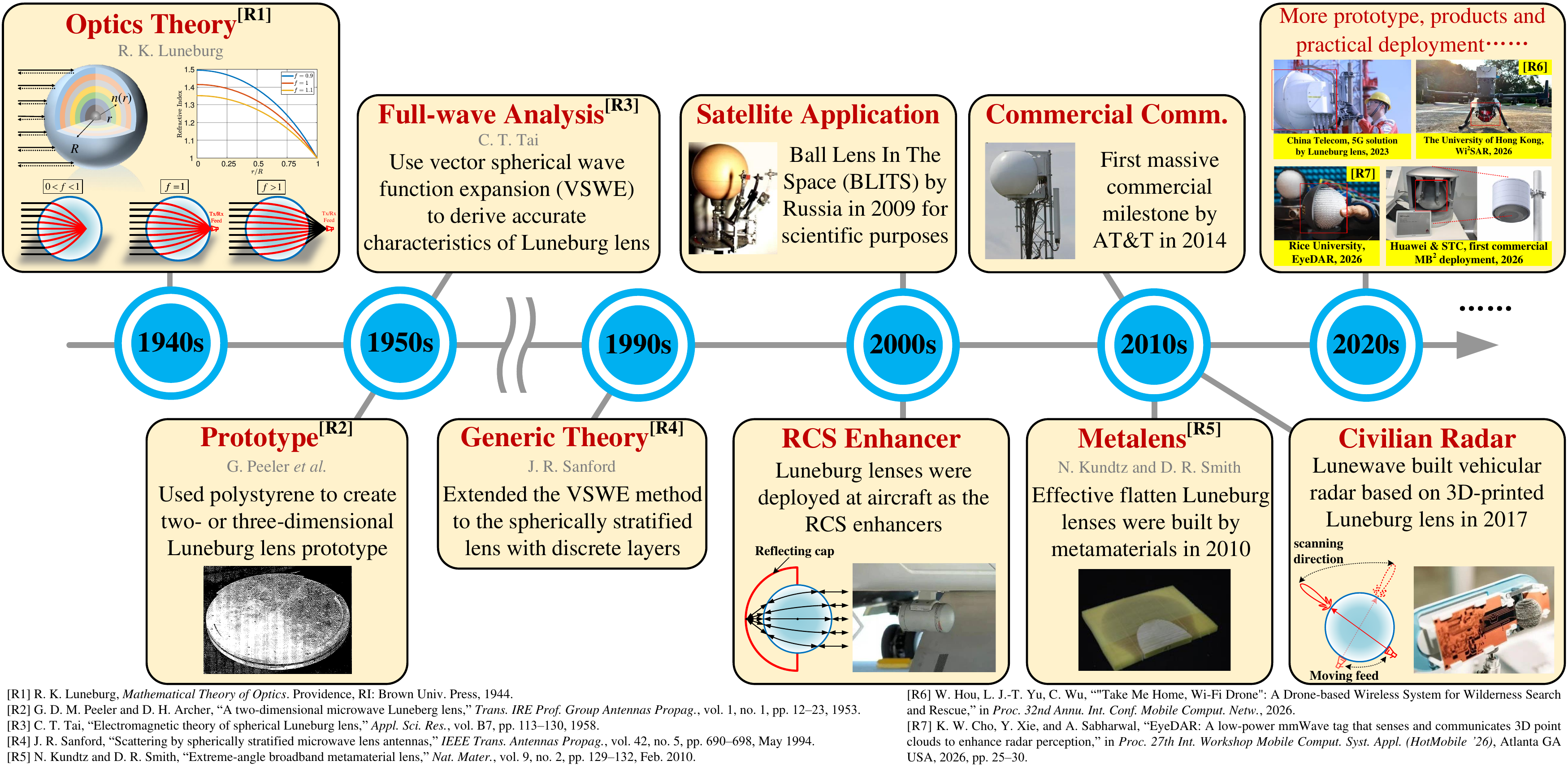}
\caption{Development timeline of Luneburg lenses. The photos are included for illustrative purposes only, and the copyright belongs to the original owners.}
\label{F1}
\vspace*{-2mm}
\end{figure*}

Fig.~\ref{F1} summarizes the development of Luneburg lenses from theory to applications. In the following, we briefly introduce the governing principle and then discuss the application milestones that motivate the notion of LUNA.

\subsection{Basic Principle}\label{S2.1}

The Luneburg lens concept was originally proposed by R.~K.~Luneburg in the 1940s \cite{Original}. For such a lens, the refractive index decreases radially from the center to the surface according to
\begin{equation*}
\label{Eq1}
n\left( r \right) = \frac{{\sqrt {1 + {f^2} - {{\left( {r/R} \right)}^2}} }}{f},
\end{equation*}
where $R$ is the lens radius, $0\le r\le R$, and $f$ controls the focal distance. Under geometric optics, a plane wave arriving from an arbitrary far-field direction is bent toward the corresponding focal point on the shadowed side, at a distance $fR$ from the center. Thus, the focus lies on the surface for $f=1$, inside the lens for $0<f<1$, and outside it for $f>1$. By reciprocity, a low-gain feed placed at that focal point is transformed into a collimated beam in the opposite direction. This one-to-one angular-to-spatial mapping, rather than active phase shifting, is the essential beam-steering mechanism of a Luneburg lens.

The subsequent theoretical milestones mark the transition from ray descriptions to quantitative electromagnetic analysis, as presented in Fig.~\ref{F1} with landmark references.
These studies established a thorough electromagnetic understanding of the focusing mechanism of Luneburg lenses and laid the theoretical groundwork for the practical lens designs.

\subsection{Applications}\label{S2.2}

The application milestones of Luneburg lenses indicate an evolution from radar uses to communication-oriented implementations. Representative applications are presented in Fig.~\ref{F1} and discussed below.

\begin{itemize}

\item {\bfseries Radar applications}: Luneburg lenses have supported radar applications since their inception. Given their focusing property, an incident wave can be focused, reflected, and re-collimated back toward the illuminator by a reflecting cap. This property makes them wide-angle radar-cross-section (RCS) enhancers.
Since the early 2000s, some aircraft have used such RCS enhancers to increase their radar visibility when required.
The same principle enabled Russia's 2009 Ball Lens In The Space (BLITS) satellite to reflect ground-transmitted signals for scientific data collection.
Another advantage of Luneburg lenses is that a feed can be moved around the focal surface to enable rapid and equal-gain beam scanning. 
In the 2010s, Luneburg lenses were increasingly employed in civilian radars. For example, Lunewave Inc. uses 3D-printed Luneburg lenses to provide the long-range, high-resolution detection required for vehicular applications \cite{Lunewave}.

\item {\bfseries Wireless communication applications}: Besides radar applications, Luneburg lenses have been adopted in wireless communication systems. Owing to the spherical symmetry, multiple feeds can be arranged on the focal surface of a single lens. Each feed corresponds to a specific beam direction, such that multiple directional beams with wide coverage and similar gain can be generated by feed switching or simultaneous feed excitation. This is attractive for multi-beam satellite links, fixed wireless access, cellular sector coverage, and backhaul/fronthaul links, where a large aperture and flexible angular coverage are needed. In 2014, AT\&T Inc. achieved an early commercial deployment of Luneburg-lens antennas \cite{MatSing}. 
Another example of high-speed railway communication supported by Luneburg lenses has been reported in \cite[Appendix III]{ITU_L1390}, demonstrating their coverage capability.

\end{itemize}

Since the early 2020s, more prototype, products, and practical deployment related to Luneburg lenses for communication and sensing applications have been reported (see the top right corner of Fig.~\ref{F1}).
These examples further confirm the potential of Luneburg lenses for future 6G-and-advanced wireless communications.

Fig.~\ref{F1} also demonstrated how manufacturing technology of Luneburg lenses mature in parallel with their applications. The earliest 1950s prototypes approximated the gradient refractive-index profile using discrete polystyrene regions. A pivotal advance came up in the 2010s, when metamaterials and transformation optics (TO) flattened the spherical geometry into a low-profile lens with a planar focal surface, greatly easing integration.
More recently, 3D printing \cite{Lunewave} has made Luneburg lenses practical and affordable at scale. Looking ahead, low-density materials and advanced fabrication are key to lightweight lens antennas for weight- and space-constrained platforms \cite{Proceeding}, where LUNA's passive front end must coexist with stringent constraints.

\section{LUNA Transceiver Architectures}\label{S3}

The development route of the Luneburg lens motivates the proposal of innovative LUNA architectures. LUNA inherits the passive focusing and high-gain beamforming capability of the Luneburg lens, while a dense bank of passive feeds and a reconfigurable feed-selection network are introduced to provide adaptive beamforming for communication, coverage extension, and sensing.
This section details the proposed LUNA architectures and analyzes their advantages over existing schemes.

\begin{figure}[t]
\captionsetup{singlelinecheck = off, font={footnotesize}, name = {Fig.}, labelsep = period}
\centering	
\includegraphics[width=0.99\columnwidth]{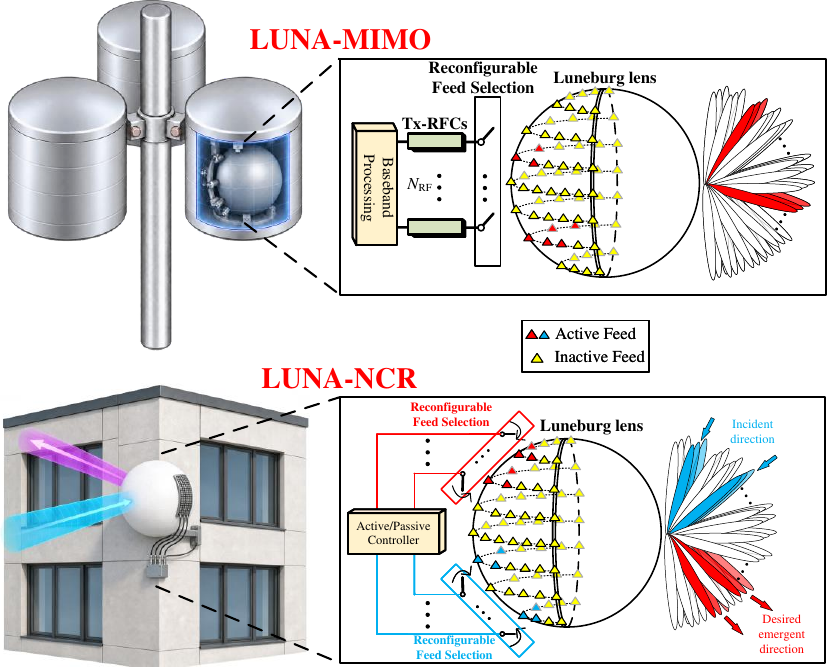}
\caption{Proposed LUNA architectures, including LUNA-MIMO and LUNA-NCR.}
\label{F2}
\vspace*{-2mm}
\end{figure}

\subsection{MIMO and NCR Meet LUNA}\label{S3.1}

We introduce LUNA for both MIMO transceivers and NCR-assisted coverage extension, as illustrated in Fig.~\ref{F2}. Note that a reconfigurable intelligent surface (RIS) can be conceptually interpreted as NCR Type-II \cite{NCRintro}, which is a passive or semi-passive variant of NCR, to align with standardized terminology. We retain the term RIS throughout this article to distinguish passive signal redirection from active amplify-and-forward relaying.

\begin{itemize}
\item {\bfseries{The LUNA-based MIMO}} (LUNA-MIMO) comprises a Luneburg lens whose focal surface is equipped with a large number of fixed, passive feeds. Unlike the active antennas used in traditional MIMO, these feeds require no dedicated power amplifiers, phase-shifter network (PSN), or RFCs. In the transmit mode, a reconfigurable feed-selection network connects only a small subset of these feeds to the available RFCs and the baseband processor, and it allows the lens to passively transform its radiation into a collimated beam in the opposite direction. Conversely, an incoming plane wave is focused onto the feed associated with its angle of arrival, and the corresponding port can then be selected for reception.
{Note that Fig.~\ref{F2} only gives the LUNA-MIMO transmitter architecture with transmit RFCs (Tx-RFCs), while the LUNA-MIMO receiver for uplink multi-user access will be detailed in sequel.}
In a nutshell, LUNA-MIMO preserves low-power passive focusing while avoiding the bulky PSN of hybrid MIMO and the one-RFC-per-antenna requirement of fully digital MIMO.

\item {\bfseries{The LUNA-based NCR}} (LUNA-NCR) uses reconfigurable feed selection to establish a controllable path between the focal feed associated with the incident direction and another feed associated with the desired emergent direction, thereby extending service coverage.
LUNA-NCR supports active and passive implementations, depending on whether amplify-and-forward circuitry is inserted into the signal path.
In the passive implementation, i.e., LUNA-RIS, the selected feeds are interconnected through a low-power switching, coupling, and phase-control network without active RF amplification. In the active implementation, an amplification chain is inserted between the selected input and output feeds to compensate for propagation and insertion losses. 
Given that high-frequency channels are usually sparse, only a few feed pairs typically suffice to map an incident direction to a desired emergent direction.
Note that in LUNA-NCR, reconfigurability is achieved through simple feed selection rather than by designing large-scale active \cite{NCRintro} or passive \cite{MyTcom} beamforming weight vectors.
\end{itemize}


\begin{table*}[t]
\centering
\captionsetup{font={footnotesize},labelsep=newline}
\caption{\sc {Comparison between the Proposed LUNA Architectures and Existing MIMO Architectures}}
\label{TAB1}
\renewcommand{\arraystretch}{1.12}
\resizebox{\textwidth}{!}{\begin{tabular}{|l|c|c|c|c|}
\hline
\diagbox{\bf Architectures}{\bf Properties} & \textbf{\begin{tabular}[c]{@{}c@{}}Power\\ Consumption\end{tabular}} & \textbf{\begin{tabular}[c]{@{}c@{}}Hardware\\ Multiplexing\end{tabular}} & \textbf{\begin{tabular}[c]{@{}c@{}}Beamforming Gain\end{tabular}} & \textbf{{Coverage}} \\ \hline

\textbf{{Proposed LUNA-MIMO}} &
\begin{tabular}[c]{@{}c@{}}{\bf Low}\\ (Passive lens with\\ few selected RFCs)\end{tabular} &
\begin{tabular}[c]{@{}c@{}}{\bf Easy}\\{(Shared spherical lens}\\ {for wideband reuse)}\end{tabular} &
\begin{tabular}[c]{@{}c@{}}{\bf High}\\ {(Equal gain; no beam}\\ {squint/edge effect)}\end{tabular} &
\begin{tabular}[c]{@{}c@{}}{\bf Wide}\\ {(Wide angle and}\\ {bandwidth coverage)}\end{tabular} \\ \hline

\textbf{{Fully Digital MIMO}} &
\begin{tabular}[c]{@{}c@{}}{Very High}\\{(One RFC per antenna)}\end{tabular} &
\multirow[c]{3}{*}{\begin{tabular}[c]{@{}c@{}}\\ \\{Difficult}\\{(Distinct architectures)}\end{tabular}} &
\multirow[c]{2}{*}{\begin{tabular}[c]{@{}c@{}}\\ {Moderate$\sim$High}\\ {(Scales with active aperture)}\end{tabular}} &
\multirow[c]{4}{*}{\begin{tabular}[c]{@{}c@{}}\\ \\ \\ \\Low\\ (Orientation-limited)\end{tabular}} \\ \cline{1-2}

\textbf{{Hybrid MIMO}} &
\begin{tabular}[c]{@{}c@{}}{High}\\ {(Large-scale PSN)}\end{tabular} &
& & \\ \cline{1-2}\cline{4-4}

\textbf{{Tri-hybrid MIMO \cite{TriHy}}} &
\begin{tabular}[c]{@{}c@{}}{Low$\sim$Moderate}\\ (Fewer antennas\\ are required)\end{tabular} &
&
\begin{tabular}[c]{@{}c@{}}{High}\\ (Extra electromagnetic\\ precoding)\end{tabular} &
\\ \cline{1-4}

\textbf{Planar Lens Array \cite{PDMA}} &
\begin{tabular}[c]{@{}c@{}}Low\end{tabular} &
\begin{tabular}[c]{@{}c@{}}Moderate\\ (Shared lens part but \\ limited frequency reuse)\end{tabular} &
\begin{tabular}[c]{@{}c@{}}Moderate\\ (Beam squint and\\ edge effects)\end{tabular} &
\\ \hline

\textbf{{MA} \cite{zhu2026tutorial}} &
\begin{tabular}[c]{@{}c@{}}{Low$\sim$Moderate}\\ {(Moving array}\\ {may be applied)}\end{tabular} &
\begin{tabular}[c]{@{}c@{}}{Difficult}\\ {(Distinct architectures)}\end{tabular} &
\begin{tabular}[c]{@{}c@{}}{Very low$\sim$Low}\\ {(Poor aperture utilization)}\end{tabular} &
\begin{tabular}[c]{@{}c@{}}{Low$\sim$Moderate}\\ (Rotation operator\\ may be applied)\end{tabular} \\ \hline

\textbf{{SIM \cite{SIM}}} &
\begin{tabular}[c]{@{}c@{}}{Moderate}\\ {(Control circuits for}\\ {passive meta-atoms)}\end{tabular} &
\begin{tabular}[c]{@{}c@{}}{Moderate}\\ {(Can be treated}\\ {as the radome)}\end{tabular} &
\begin{tabular}[c]{@{}c@{}}{Moderate}\\ {(High DoF but with}\\ {cascading loss)}\end{tabular} &
\begin{tabular}[c]{@{}c@{}}Low\\ (Orientation-limited)\end{tabular} \\ \hline
\end{tabular}%
}
\par\vspace{2mm}

\renewcommand{\arraystretch}{1.12}
\resizebox{\textwidth}{!}{\begin{tabular}{|l|c|c|c|c|}
\hline
\diagbox{\bf Architectures}{\bf Properties} & \textbf{\begin{tabular}[c]{@{}c@{}}Power\\ Consumption\end{tabular}} & \textbf{\begin{tabular}[c]{@{}c@{}}Hardware\\ Multiplexing\end{tabular}} & \textbf{\begin{tabular}[c]{@{}c@{}}Beamforming Gain\end{tabular}} & \textbf{{Coverage}} \\ \hline

\begin{tabular}[c]{@{}l@{}}{\bf Proposed LUNA-NCR}\end{tabular} &
\begin{tabular}[c]{@{}c@{}}{\bf Low $\sim$Moderate} \\ {(passive or active)}\end{tabular} &
\begin{tabular}[c]{@{}c@{}}{\bf Easy}\\ {(Shared spherical lens}\\ {for wideband reuse)}\end{tabular} &
\begin{tabular}[c]{@{}c@{}}{\bf High}\\ (Equal gain; no beam \\ squint/edge effect)\end{tabular} &
\begin{tabular}[c]{@{}c@{}}{\bf Wide}\\ {(Wide angle and}\\ {bandwidth coverage)}\end{tabular} \\ \hline

\textbf{{Standardized NCR} \cite{NCRintro}} &
\begin{tabular}[c]{@{}c@{}}{High}\\ {(Active amplification}\\ {and dual analog precoding)}\end{tabular} &
\multirow[c]{2}{*}{\begin{tabular}[c]{@{}c@{}} \\{Difficult}\\ {(Distinct architectures)}\end{tabular}} &
\begin{tabular}[c]{@{}c@{}}{Moderate$\sim$High}\\ {(Scales with active aperture)}\end{tabular} &
{\begin{tabular}[c]{@{}c@{}}Low\\ (Orientation-limited)\end{tabular}} \\ \cline{1-2}\cline{4-5}

\begin{tabular}[c]{@{}l@{}}{\bf Conceptual NCR Type-II}\\ {(i.e., RIS \cite{MyTcom})}\end{tabular} &
\begin{tabular}[c]{@{}c@{}}Low$\sim$Moderate\\ (Few active elements \\ may be applied)\end{tabular} &
&
\begin{tabular}[c]{@{}c@{}}High\\ (Massive elements)\end{tabular} &
{\begin{tabular}[c]{@{}c@{}}Low$\sim$Moderate\\ (Transparent RIS \\ may be applied)\end{tabular}} \\ \hline
\end{tabular}%
}
\vspace*{-2mm}
\end{table*}

\subsection{Advantage Analysis}\label{S3.2}

To highlight the technological features of the proposed LUNA scheme, Table~I compares the LUNA architectures with state-of-the-art counterparts. The upper part compares LUNA-MIMO, in sequence, with fully digital MIMO, hybrid MIMO, tri-hybrid MIMO \cite{TriHy}, planar lens arrays \cite{PDMA}, MAs \cite{zhu2026tutorial}, and stacked intelligent metasurfaces (SIMs) \cite{SIM}. The lower part compares LUNA-NCR with its conventional version (including RIS as NCR Type-II).

In terms of power consumption, fully digital MIMO assigns an RFC to each antenna and therefore is the most power-hungry, whereas hybrid MIMO reduces RFCs via an active PSN, but remains power-intensive. Tri-hybrid MIMO adds antenna-domain precoding through pattern-reconfigurable antennas, requiring extra control and joint optimization across multiple domains.
A SIM performs wave-domain precoding with stacked passive programmable meta-atoms, lowering RF-related energy consumption yet inducing extra bias/control circuits and cumulative insertion loss.
LUNA-MIMO, however, differs from the above architectures by using a single passive lens and a reconfigurable network only to select the required focal feeds.
For LUNA-MIMO, only a small subset of feeds is connected to a few RFCs for given tasks. 
For LUNA-NCR, the passive LUNA-RIS implementation retains the low-power nature of the lens and feeds, while the active implementation replaces dual analog beamforming with simple feed selection.

The Luneburg lens constitutes a common electromagnetic front end that is reused rather than rebuilt per service. Functionally, the same lens operates as LUNA-MIMO when selected feeds connect to RFCs and the baseband, and as LUNA-NCR when selected feed pairs are interconnected through passive or active paths. The lens also supports cross-band reuse. Note that 5G New Radio defines frequency ranges (FRs) of $410$\,MHz--$7.125$\,GHz (FR1) and $24.25$--$71$\,GHz (FR2), and 6G is expected to add FR3 ($7.125$\,GHz to $24.25$\,GHz) and sub-terahertz bands. A single Luneburg lens can serve multiple bands without beam squint, while most counterparts lack hardware that is easily multiplexed across services and bands.

On beamforming gain, MAs have limited aperture utilization and therefore poor beamforming gain, while schemes based on planar arrays suffer from edge effects and beam squint at large scan angles.
A SIM can realize relatively high gain but relies heavily on multilayer meta-atoms and fabrication accuracy.
Thanks to its spherical symmetry, the Luneburg lens provides an approximately angle-invariant focusing response over a large angular spread, and the low-cost, low-complexity feed selection of LUNA exploits this response for equal-gain, wide-angle beamforming without edge effects or beam squint.
Moreover, the dense feed bank provides an opportunity for uplink multiple access. Signals from users in different directions are mapped onto spatially separated focal regions, where the feeds can separate users sharing the same time-frequency resource.

As for coverage, planar lens arrays, SIMs, fully digital/hybrid/tri-hybrid MIMO arrays, and traditional RISs are sensitive to aperture orientation and experience coverage degradation at large scan angles. The MAs equipped with rotation mechanism can improve coverage at the cost of additional mechanical control.
Conventional NCRs mainly serves predefined donor and service sectors, and it has to sacrifice field of view for self-interference suppression. 
LUNA, by contrast, supports wide-angle transmission and reception with approximately equal gain through reconfigurable feed selection alone.

\subsection{Vision and Potential Applications of LUNA}\label{S3.C}

\begin{figure*}[!tp]
\centering
\captionsetup{singlelinecheck = off, font={footnotesize}, name = {Fig.}, labelsep = period}
\subfigure{\includegraphics[width=1.98\columnwidth]{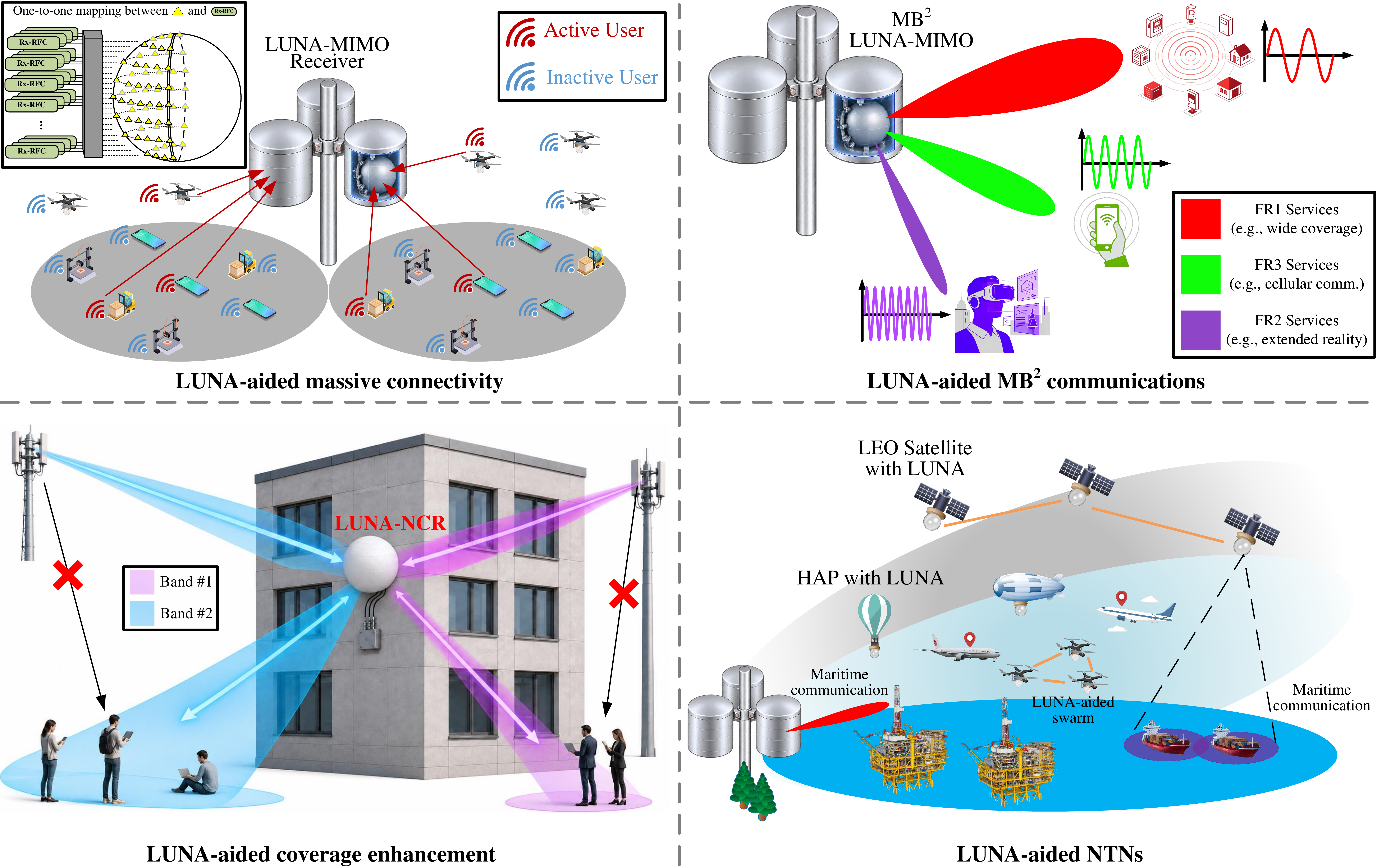}}
\caption{Potential use cases supported by the proposed {LUNA} architectures.}
\label{F3}
\vspace*{-2mm}
\end{figure*}

We focus on four representative 6G applications that can directly exploit the passive large-aperture focusing, wide-angle feed selection, and hardware-reuse capabilities of LUNA, as illustrated in Fig.~\ref{F3}.

\begin{itemize}

\item \textbf{Massive connectivity}:
A dedicated LUNA-MIMO receiver architecture can support massive connectivity in 6G. A key feature is the one-to-one mapping between the passive feeds and dedicated receive RFCs (Rx-RFCs), enabling the lens to capture signals arriving from various directions simultaneously. Compared with Tx-RFCs, Rx-RFCs generally consume substantially less power because they do not require high-power amplifiers, while the feeds themselves remain passive. The dense feed bank samples the spherical focal surface so that signals from different directions are focused onto separated focal regions and routed to the corresponding Rx-RFCs. Multiple active users can therefore be separated in the beamspace domain while sharing the same time-frequency resource block, supporting scalable access for diverse equipment types.

\item \textbf{Multi-band multi-beam (MB$^2$) communications}:
LUNA naturally supports MB$^2$ communications by concurrently activating different feeds to form independently directed beams for heterogeneous services. As shown in Fig.~\ref{F3}, one LUNA-MIMO aperture can simultaneously provide FR1 services (e.g., wide-area coverage), FR2 services (e.g., extended reality), and the upcoming FR3 services in 6G. The spherical lens offers frequency-invariant spatial focusing, allowing the beams to operate over these bands with reduced beam-squint-induced misalignment. Sharing one lens and feed bank across FR1--FR3 reduces the antenna count, tower space, and installation cost of co-located multi-band arrays, while feed selection allows the beams and associated RFCs to be reassigned as service demands change.

\item \textbf{Coverage enhancement}:
LUNA-NCR can provide an alternative link when the direct line-of-sight paths from base stations (BSs) to their user groups are blocked. The donor-direction feed and a service-direction feed are selected on the same spherical lens, enabling the incident signals to be redirected toward the obstructed or cell-edge regions. The active implementation inserts amplification, whereas the passive LUNA-RIS redirects signals through low-power switching and coupling. A key advantage of LUNA-NCR over conventional NCR and RIS is that the same lens and feed bank can be readily reused to form multiple beams over different frequency bands, making MB$^2$ coverage extension practical and broadening the deployment scenarios of NCR-assisted services.

\item \textbf{Non-terrestrial networks (NTNs)}:
LUNA is promising for unmanned aerial vehicle (UAV) swarms, high-altitude platforms (HAPs), low-Earth-orbit (LEO) satellites, and maritime communication platforms, as depicted in Fig.~\ref{F3}. These systems require high-gain links over long distances while their relative geometries vary rapidly. Electronic feed selection over the spherical focal surface can support wide-angle beam steering and tracking without mechanical gimbals or multiple orientation-specific planar arrays. Compared with a conventional planar aperture, the smooth spherical profile also presents lower and less orientation-dependent wind loading, reducing wind-induced torque, vibration, and support loads on airborne and maritime platforms.

\end{itemize}

\section{Numerical Analysis for Case Studies}\label{S4}

\begin{figure}[t]
\captionsetup{singlelinecheck = off, font={footnotesize}, name = {Fig.}, labelsep = period}
\centering	
\includegraphics[width=0.98\columnwidth]{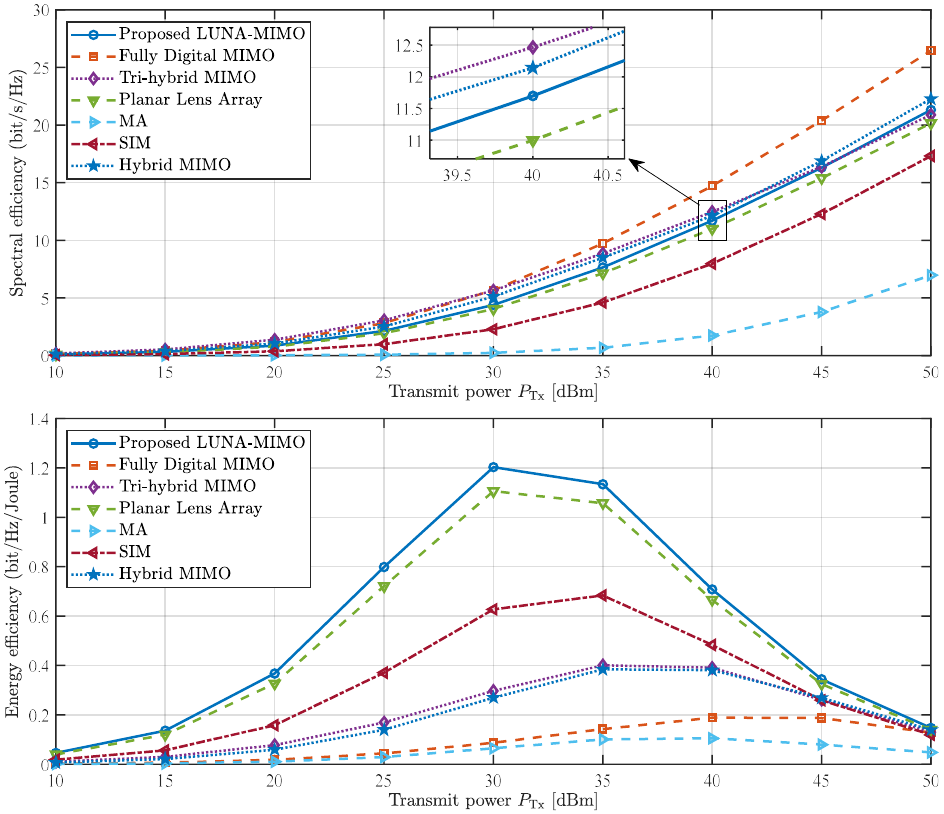}
\caption{Comparison of the spectral efficiency (upper) and energy efficiency (lower) between LUNA-MIMO and the existing state-of-the-art.}
\label{F4}
\vspace*{-2mm}
\end{figure}

We evaluate the proposed LUNA architectures through case studies, comparing communication and sensing performance under consistent fairness rules. We consider a cellular network operating at carrier frequency $f_{\rm c} = 30$\,GHz (corresponding wavelength $\lambda \approx 10$\,mm).
Orthogonal frequency division multiplexing (OFDM) with $32$ subcarriers and bandwidth $500$\,MHz is adopted.
For comparison, we consider the uniform planar array (UPA) MIMO with fully digital and hybrid architectures, {MA} \cite{zhu2026tutorial}, planar lens array \cite{PDMA}, SIM \cite{SIM}, and tri-hybrid MIMO \cite{TriHy}.
The radius of the adopted Luneburg lens is set to $R = 4.51\lambda$, which yields physical aperture $\pi R^2 \approx 64\lambda^2$. For fairness, the considered fully digital/hybrid/tri-hybrid MIMO and SIM are assumed to have $16 \times 16$ half-wavelength spaced antennas, and the MA and planar lens array are of the same physical size $8\lambda \times 8\lambda$. The number of movable elements at the MA is set to $4$.
Except for the fully digital MIMO with $256$ RFCs, all other schemes have $4$ RFCs.

Fig.~\ref{F4} compares spectral efficiency (SE) and energy efficiency (EE) versus the transmit power, with EE defined as the SE per unit of total consumed power \cite{Sun}.
Fully digital MIMO with one RFC for each antenna sets the SE upper bound, but at the cost of overwhelming $256$ RFCs, yielding the lowest EE.
The central result is that LUNA-MIMO approaches the hybrid and tri-hybrid benchmarks using only $4$ RFCs and a low-complexity switching network. Note that the considered PSN contains $1024$ phase shifters. Because the Luneburg lens performs broadband focusing without a large PSN, LUNA-MIMO attains the highest EE among all schemes, delivering the quantitative embodiment of its design proposition.

Given the OFDM system, architectures, such as hybrid MIMO, tri-hybrid MIMO, and SIM, whose analog processing relies on frequency-independent PSN, cannot simultaneously realize their optimum beamforming responses over all subcarriers, which limits their SE performance. 
The planar lens array can adopt low-complexity feed selection and thus shows favorable EE, yet its frequency-dependent beam directions incur a significant loss compared with LUNA-MIMO.
The MA, with only a few active elements, attains the lowest SE and gains little EE benefit from its limited aperture.
LUNA-MIMO avoids the wideband penalty, which is a distinct advantage under the large bandwidths envisioned for 6G.

\begin{figure}[t]
\captionsetup{singlelinecheck = off, font={footnotesize}, name = {Fig.}, labelsep = period}
\centering	
\includegraphics[width=0.99\columnwidth]{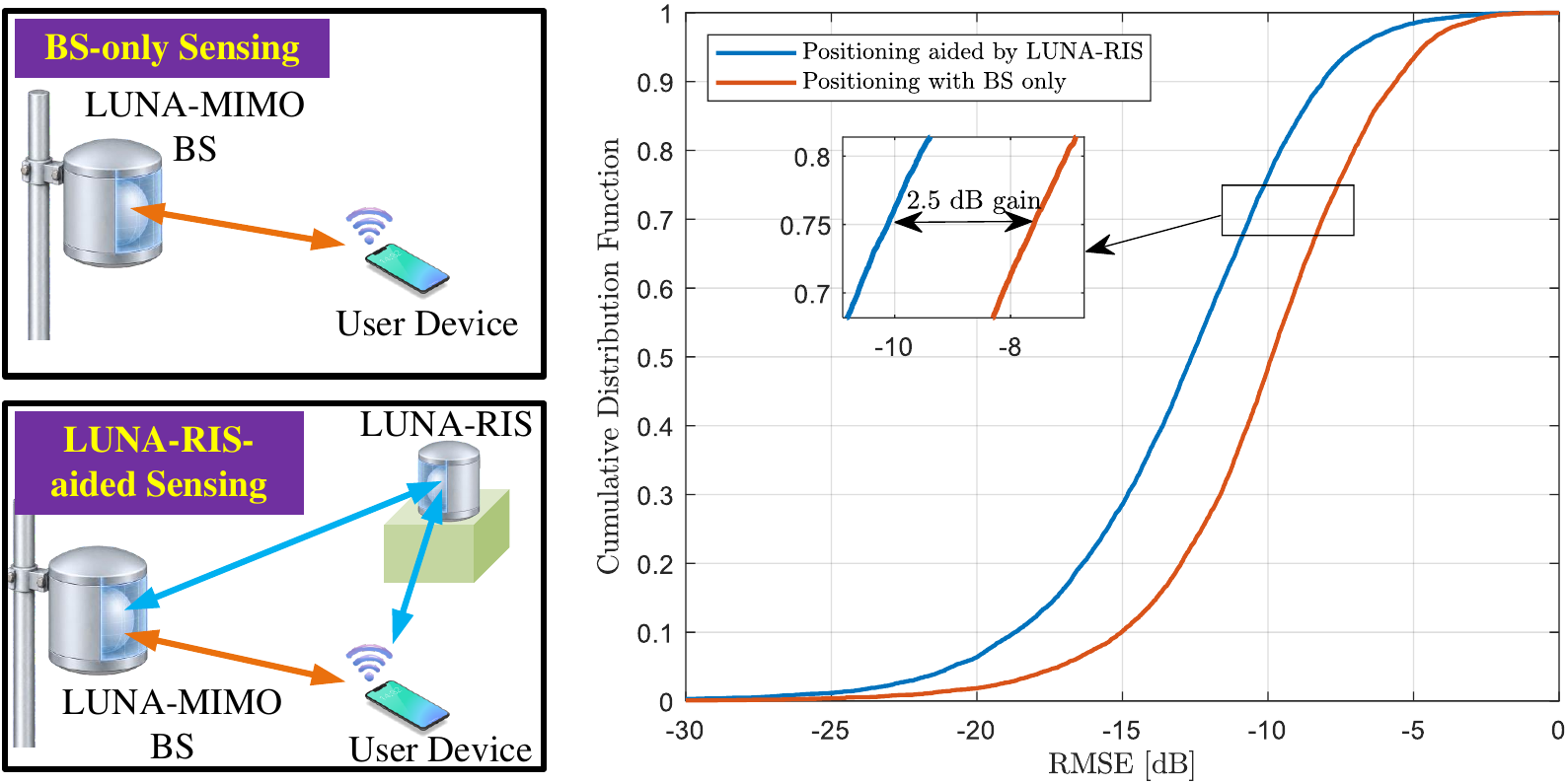}
\caption{Positioning performance with the aid of the proposed {LUNA-MIMO}/{LUNA-RIS}.}
\label{F5}
\vspace*{-2mm}
\end{figure}

Furthermore, the sensing performance assisted by LUNA-MIMO and/or LUNA-RIS is evaluated.
Fig.~\ref{F5} quantifies the positioning performance in terms of the root mean-square error (RMSE) between the actual position and estimated position.
Observe that we can attain submeter-level positioning (RMSE $< 0$ dB) upon using the proposed LUNA paradigm.
Moreover, with the LUNA-RIS whose position is known, the RIS-aided sensing can be viewed as a virtual multistatic radar system. In this sensing system, therefore, the user's positions relative to both the BS and the RIS are estimated and then jointly leveraged for more accurate positioning.
One can see from Fig.~\ref{F5} that compared to the BS-only sensing, the positioning performance with the LUNA-RIS achieves $2.5$ dB RMSE gain for $76\%$ of cases.
This verifies the effectiveness of the proposed {LUNA} architecture in sensing tasks.

\section{Open Problems and Future Research Directions}\label{S5}

This section outlines some open problems and research directions for the proposed LUNA paradigm that require further investigation.

\subsection{Theoretical Analysis}\label{S5.1}

Introducing a Luneburg lens and reconfigurable selection over a dense passive feed bank changes both electromagnetic (EM) propagation and the effective beamspace channel. Geometric-optics and ray-tracing models describe angular focusing but cannot fully capture mutual coupling, diffraction, focal-spot distortion, and material loss. EM signal and information theory (ESIT) \cite{ESIT}, which connects Maxwell's equations with Shannon theory, is therefore needed to establish tractable channel models and performance bounds for LUNA-MIMO and LUNA-NCR. Key problems include characterizing the DoFs, capacity, focusing efficiency, and angular resolution under discrete feed selection and limited RFC connectivity. These results should quantify how lens size, feed density, port coupling, and hardware impairments jointly determine performance and guide architecture design and resource allocation.

\subsection{Hardware Design}\label{S5.2}

Practical LUNA implementation requires a precise, lightweight, and broadband lens together with a scalable feed-selection network. Conventional 3D dielectric Luneburg lenses are bulky and heavy, restricting their use on space- and weight-constrained platforms. Promising solutions include low-density graded-index materials, hollow or discretized spherical structures, additive manufacturing, and integrated feed supports that retain spherical symmetry while reducing loss and weight. Over the focal surface, high-speed switch matrices based on PIN diodes, radio-frequency microelectromechanical systems, or pixel-antenna-inspired interconnections \cite{Keke} may connect only the selected feeds to the RFCs. Their joint design must balance feed density, port isolation, insertion loss, switching speed, control power, thermal behavior, and environmental reliability, followed by validation using representative hardware prototypes.

\subsection{Near-Field LUNA Transmission}\label{S5.3}

When the LUNA aperture becomes electrically large relative to the link distance, spherical-wave propagation must be considered. Unlike a far-field plane wave, a near-field wave changes both the angular and radial location of the focal region. As the source approaches the lens, the focus may move beyond the physical focal sphere ($r>R$). Near-field LUNA may therefore require a multilayer, conformal, or otherwise tunable feed structure that supports joint angular and radial selection. This additional distance-angle coupling can also enable user separation, localization, and spatial focusing beyond conventional angular beamforming. Important research topics include near-field channel estimation, feed selection and precoding, spherical-wave codebook design, transition-region modeling, and communication-sensing trade-offs under finite feed resolution, mutual coupling, and limited RFCs.

\subsection{AI-Driven Solutions for LUNA}\label{S5.4}

AI can coordinate channel estimation, channel-state-information feedback, feed selection, beam tracking, and mode switching among LUNA-MIMO, LUNA-RIS, and LUNA-NCR. Model-driven learning may exploit the structured beamspace channel to reduce training data and computational cost, while improving robustness to model mismatch, low-resolution analog-to-digital/digital-to-analog converters, power-amplifier nonlinearity, and in-phase/quadrature imbalance. However, practical deployment must address generalization across lens sizes, frequency bands, and propagation environments, together with inference latency, control overhead, and energy consumption at the network edge. Digital twins, self-supervised adaptation, and lightweight on-device learning are promising directions for closed-loop configuration, provided that interpretability and reliability can be maintained for safety-critical communication and sensing services.

\section{Conclusions}\label{S6}

This article has introduced LUNA, a reconfigurable array architecture that combines the passive focusing capability of a Luneburg lens with a dense passive feed bank and reconfigurable feed selection. We have developed LUNA-MIMO and LUNA-NCR, with the latter supporting active amplify-and-forward and passive LUNA-RIS implementations through a shared lens-and-feed front end. LUNA enables high-gain, wide-angle transmission, reception, and coverage extension with reduced hardware complexity, while facilitating hardware reuse across functions and frequency bands. Representative applications have demonstrated its broad relevance to future wireless networks, and numerical case studies have illustrated its potential benefits for communications and sensing.
In essence, we believe that LUNA establishes a new paradigm shift for 6G-and-advanced networks and this article would stimulate further related research.

\balance


\begin{thebibliography}{99}

\bibitem{itu2023framwork} 
ITU-R, ``M.2160: Framework and overall objectives of the future development of IMT for 2030 and beyond,'' 2023. [Online]. Available: https://www.itu.int/rec/R-REC-M.2160/en

\bibitem{zhu2026tutorial} 
L. Zhu {\it et al.}, ``A tutorial on movable antennas for wireless networks,''
{\it IEEE Commun. Surv. Tutor.}, vol.~28, pp.~3002--3054, 2026.

\bibitem{Original} 
R.~K.~Luneburg, {\it Mathematical Theory of Optics}. Providence, RI: Brown Univ. Press, 1944.







\bibitem{Lunewave} 
Lunewave Inc., ``Automotive radar sensor,'' [Online]. Available: \url{ https://lunewave.com/automotive-radar-sensor/}.

\bibitem{MatSing} 
P. Afanasyev {\it et al.}, ``Multi-beam Luneburg lens antenna for cellular communications,''
in {\it Proc. 9th Eur. Conf. Antennas Prop. (EuCAP)}, Apr. 2015, pp.~1--4.

\bibitem{ITU_L1390} 
ITU-T, ``Energy saving technologies and best practices for 5G radio access network (RAN) equipment,'' \emph{ITU-T Recommendation L.1390}, Aug. 2022.

\bibitem{Proceeding} 
Z. N. Chen {\it et al.}, ``Microwave metalens antennas,''
{\it Proc. IEEE}, vol.~111, no.~8, pp.~978--1010, Aug.~2023.

\bibitem{NCRintro} 
Y. Zhao {\it et al.}, ``Evolution from NCR for 5G-Advanced to RIS for 6G: A unified system framework and standardization roadmap,''
{\it IEEE Commun. Standards Mag.}, doi: 10.1109/MCOMSTD.2026.3718533.

\bibitem{MyTcom} 
S. Guo {\it et al.}, 
``Reconfigurable intelligent surface empowered simultaneous communication, localization, and mapping for vertical applications,''
{\it IEEE Wireless Commun.}, vol.~32, no.~5, pp.~134--141, Oct. 2025.

\bibitem{TriHy} 
P. Zheng, Y. Zhang, T. Y. Al-Naffouri, M. J. Hossain and A. Chaaban, ``Tri-hybrid multi-user precoding using pattern-reconfigurable antennas: Fundamental models and practical algorithms,''
{\it IEEE Trans. Commun.}, vol.~74, pp.~11789--11804, 2026.

\bibitem{PDMA} 
Y.~Zeng and R.~Zhang, 
``Millimeter wave MIMO with lens antenna array: A new path division multiplexing paradigm,''
{\it IEEE Trans. Commun.}, vol.~64, no.~4, pp.~1557--1571, Apr. 2016.

\bibitem{SIM} 
J. An {\it et al.}, ``Stacked intelligent metasurface-aided MIMO transceiver design,''
{\it IEEE Wireless Commun.}, vol.~31, no.~4, pp.~123--131, Aug. 2024.

\bibitem{Sun} 
Y. Sun {\it et al.}, ``Principal component analysis-based broadband
hybrid precoding for millimeter-wave massive MIMO systems,''
{\it IEEE Trans. Wireless Commun.}, vol.~19, no.~10, pp.~6331--6346, Oct. 2020.

\bibitem{ESIT} 
M. D. Renzo and M. D. Migliore, ``Electromagnetic signal and information theory,''
{\it IEEE BITS Inf. Theory Mag.}, vol.~4, no.~1, pp.~25--39, Mar. 2024.

\bibitem{Keke} 
K. Ying {\it et al.}, ``Reconfigurable massive MIMO: Harnessing the power of the electromagnetic domain for enhanced information transfer,''
{\it IEEE Wireless Commun.}, vol.~31, no.~3, pp.~125--132, Jun. 2024.


\end{thebibliography}
\end{document}